# The Physicists and the Historians of Physics

Pasquale Tucci[1]

*Abstract*: In the 60s of the last century the few courses of History of physics in Physics degree were held by scholars who, apart from a few exceptions, did not have a specific research background in the field. Some activities, books, social movements in the civil society allowed in the 70's the entry, among Physics courses, of teachings in History of physics held by scholars specifically trained for that job since their degree. A second change happened in the 90s when many difficulties forced physicists to allocate fewer and fewer resources to the History of their discipline. I'll outline the features of the two periods and the efforts of Historians to find a proper space in Physics departments.

## 1. Introduction

In the 1985 Silvio Bergia took on the task of drawing up an assessment of National Group of History of Physics (GNSF) activities and of papers published in the Proceedings of the yearly conferences (Bergia 1985). Fabio Bevilacqua and Salvatore Esposito outlined the historical context of the birth of the GNSF, of the first meetings and of the first national conference in Pavia and, subsequently, of establishment of SISFA in Milan (Bevilacqua, Esposito 2021). Gerardo Ienna has published a contribution on the birth of researches in History of physics and astronomy (HPA) in Italian scientific institutions (Ienna 2022).
Other contributions have been published on the *status* of the History of science in Italy (Torrini 1988, Pancaldi 1980, Pancaldi 2020). From them epistemological as well as institutional break between the History of science and the History of scientific disciplines emerge. The rationale of this dichotomy is long-lasting, but I'll not address this issue. I'll concentrate my attention on social and cultural events which allowed a strong interest in physics departments[2] for researches and teachings of history in the '70s and the fall of interest in the '90s.

## 2. Physicists as Historians of physics in Italy

In Italy, in the second half of the '900, some physicists who were still active as researchers in some area of their discipline, dedicated time and effort to the History of physics. The precursor was Giovanni Polvani (1892-1970) who, already before and during the Second World War, had expressed his interest in the History of physics.[3]

---

[1] Università degli Studi di Milano (retired in 2013) - ptucci@icloud.com
[2] I'll use "department" as to denote structures with disciplinary homogeneity within larger institutions such as Universities or research organizations. An exact definition of "department" will be introduced, on an experimental basis, in the University Law of 1980.
[3] Polvani published a book on Antonio Pacinotti, a book on the nature of light, a memoir on the Italian contribution to the progress of physics, an essay on A. Volta. After the war he published, together with Luigi Gabba, a collection of writings by O.F. Mossotti. On Polvani's institutional responsibilities and on his person as physicist and historian of physics see (Gariboldi 2022a, pp. 57-62).

As President of the Italian Physical Society (SIF) from 1947 to 1961, as President of the National Research Council (CNR) from 1960 to 1965 and as President of the Domus Galilæana from 1955 to 1970 he worked so that studies in history had a recognizable and shared *status* of research and education within scientific institutions. In the1961 he was responsible for the inclusion of courses of History of physics, of Epistemology and methodology in Physics degree courses.

The other physicist who dedicated a considerable part of his life to the History of physics was Edoardo Amaldi (1908-1989). On Amaldi as a physicist and as a statesman see (Rubbia 1991), (Battimelli *et al.* 2022). In the 1966 he published a monograph on Ettore Majorana and in 1979, together with Rasetti, he published a monograph on Enrico Persico. In the 2012 a manuscript of him about the adventurous life of Friedrich Houtermans was published posthumously (Braccini *et al.* 2012). Amaldi scrupulously reconstructed the events, and built their myth, of the boys of Via Panisperna and the contribution by Roman physicists to the discovery of neutron and nuclear fission (Amaldi 1984). Without going into too many details, I would like to emphasize the importance of a work by Amaldi on *Physics Today* (Amaldi 1973). In it he highlighted not only his deep knowledge of different sectors of physics, but he historically reconstructed the path of physics starting from two Planck's lectures of the1908 and the 1929. Amaldi framed his physical and historical arguments in a philosophically committed context.

Guido Tagliaferri (1920-2000), who had been a close collaborator of Polvani, and good friend of Amaldi, combined research in Physics and in the History of physics. In 1979 he transferred his physics chair of Structure of matter to History of physics, the first in Italy (Gariboldi 2022b, pp. 163-164). Once Salvini told me, mentioning Beppo Occhialini's remarck, that "... when there is Guido Tagliaferri, Milan becomes a little Athens." (Salvini, Tucci 2000, p. 68). In 1985 Tagliaferri – *cutiron* or *cutirons* or simply *cuti*, his nickname (Salvini, Tucci 2000, p. 66; Salvetti 2002, p. 178) - published the book *History of Quantum Physics*. In it the main papers of quantum physics from Planck to wave mechanics were examined in detail. (Tagliaferri 1985; Robotti 2022) The book was the result of his lectures to physics students. But his first experience as a professor of History of physics was disappointing.

At the end of the course the students expressed their disappointment to me: they wanted a descriptive course, not what turned out to be in another physics course. ... I recognized that the method of reading the original papers required more personal work from students than they could dedicate to physics. So, in the following years, I began to write lecture notes illustrating (favoring the need for comprehensibility) the historical development of quantum as reconstructed by examining the most significant works of its founders. ... In the following years I stabilized the content of the course, giving it the form that appears in my book *History of Quantum Physics*. (Tagliaferri 1994, pp. 374-375).

Tagliaferri has given many other contributions to the History of physics and astronomy. Among them I cite (Mandrino *et al.* 1994), (Tagliaferri, Tucci 1999). He was also a promoter of initiatives for safeguarding, inventorying and cataloguing of scientific instrumentation no longer used in researches, archives of scientists and of scientific institutions, scientific libraries. The Brera Astronomical Museum was the first new museum in Lombardy to obtain regional recognition (Miotto *et al.* 1989; Tucci 2007).Tagliaferri considered his experience as a professor of History very stimulating, but he found modest the appreciation of his fellow physics professors, and in any case of little practical consequence. It was the signal of a change in physicists' attitude that would be evident a few years later. I will come back to this issue afterwards.

To conclude on physicists as historians I would like to underline how Polvani, Amaldi and Tagliaferri were not only talented physicists but had a deep historical and philosophical background. And, probably, they were just the tip of the iceberg of a widespread phenomenon among physicists of their time.

### 3. Cultural context in the 70s

The practice of researches in History of physics and in History of astronomy (SF&SA) in scientific research institutes is a rather recent phenomenon and located, temporally, in the second half of the '900. In that period physicists and astronomers, stimulated by students, often involved in protest movements active on the civil and political scene, felt the need to broaden the horizon of their researches.
A radical change took place in the 70s when the History courses begun to be assigned to young scholars who have been trained to carry out researches in SF&SA since the years of their university education.
Moreover, the courses of SF&SA were completely independent, from institutional point of view, from those of the History of science. The same happened, although to a lesser extent, in other scientific disciplines such as Chemistry, Biology, Geology. Different considerations (which I'll not address) deserve the History of mathematics and History of medicine. In this way, after more than half a century, what Aldo Mieli had hoped for was realized, namely

... that the teaching of the history of mathematics, medicine, physics, chemistry and some of the natural sciences be counted among the complementary courses recommended for applicants for a degree in science or medicine, and that in the major centers be an official course." (Mieli 1916).

Finally, it should not be overlooked that closely working with men and women of science entailed for historians of scientific disciplines sharing of a style of work and research widespread in institutions dedicated to scientific research.
Historical studies flourished in Italy not only in the scientific researches institutions but were also practiced by people staffed in institutions that do not carried out researches in history as their main task. First GNSF and, afterwards, SISFA have supported this trend. In their annual conference there were no constraints on participation of unstaffed free scholars. The uninterruptedly published Proceedings since 1983 provide tangible testimony to this.[4]
Particularity of Italian situation had consequences not only in an increase of courses of History in degree courses in Physics and Astronomy;[5] but it also meant introduction of original methodologies suitable for analyzing nineteenth or twentieth century topics and scientists' memoirs and books. The phenomenon, not new on the international scene (just think of the studies of John Louis Emil Dreyer or Stillman Drake), was quite new in the Italian panorama of the history of science where a literary and humanistic background was prevalent among historians of science and where the single sciences were subject to an all-encompassing definition of "science".

---

[4] The first Proceedings were published in 1983 and refer to the third Congress held in Palermo in 1982. In the 1981 a meeting was held in Pavia, before the first national conference. Already months before the meeting Lanfranco Belloni, Fabio Bevilacqua, Enrica Giordano, Guido Tagliaferri, Pasquale Tucci had periodically met in Milan to discuss topics of History of Physics.

[5] It is good to keep in mind, however, that courses in history, epistemology, philosophy of science were a very small subset of courses offered to students who enrolled in physics.

But thorough attention that Historians of science usually devoted to historiographical problems and to secondary literature was almost entirely lost as Historians of physics privileged historical reconstruction of memoirs and books of nineteenth and twentieth-century scientists: a not at all easy job, however, useful also for Historians of science.

To Historians of physics and astronomy this meant a partial departure from tradition of historical studies on science inaugurated, conventionally, by Sarton in 1903 with the creation, in Belgium, of the ISIS magazine.[6] It became a magazine of the History of Science Society founded in 1925 in the USA by Sarton himself and it's still published continuously. In the first issue of ISIS, Sarton remarked as what he called "analytical tendency" had been instrumental in the development of science: scientists had become more and more specialists in increasingly narrow areas. But science risked missing its goal if we reduced it only to the discovery of isolated facts and lost sight of the organization and explanation of facts in an organic vision. The consequences of such a trend not only would make scientists lose sight of the primary goal of science but would threaten social life itself:

Loin de pouvoir songer à unir les hommes par des points de vue communs, les savants finiraient par ne plus se comprendre eux mêmes. (Sarton 1913, pp. 3-4)

History of science provided, according to Sarton, the best instrument of synthesis. He emphasized that History of science had to be established as an independent discipline with its own methods and with its own working tools: manuals, bibliographies, etc.

In the Introduction to his monumental work on the *History of Science*, he stressed how development of science was an essential phase of human civilization (Sarton 1927).

If Sarton gave rise to a new era in the History of science, this didn't mean that previously History of science didn't exist. Indeed, History of science and History of single scientific disciplines had a long tradition in Europe.

Since the end of the 1700s several works of History of single sciences had been published.[7] Despite this long tradition in the History of science and in the History of physics, only after the second World War did the teaching of History of physics entered the degree courses in physics.

Alongside researches, SF&SA teachings multiplied between the 60s and the end of the 80s, driven by:

---

[6] George Sarton (1884-1956). Born and educated in Belgium, he moved to the United States in 1915, where he received citizenship in the 1924.

[7] Limiting myself to the History of physics and that of astronomy, and considering those of History of mathematics where physics and astronomy were included, I mention: Priestley, *The History and Present State of Electricity, with original experiments*; Montucla, *Histoire des Mathématiques*; La Place: "Précis de l'histoire de l'astronomie"; Delambre: *Histoire de l'Astronomie ancienne e de l'Astronomie moderne*; Höfer: *Histoire de la physique et de la chimie*; Tannery: *Recherches sur l'histoire de l'astronomie ancienne*; Poggendorff: *Geschichte der Physik*; Cajori: *A History of Physics in Its Elementary Branches*; Mach: *Die Mechanik in ihrer Entwicklung. Historisch-kritisch dargestellt*; Dreyer: *History of the planetary systems from Thales to Kepler*; Duhem: *Le système du monde* (Duhem 1913); Dijksterhuis: *The Mechanization of the world picture*.
Between the 1868 and the 1887 the *Bullettino di Bibliografia e di Storia delle Scienze Matematiche e Fisiche* (*Bulletin of Bibliography and History of Mathematical and Physical Sciences*) was edited by Baldassarre Boncompagni. The *Bullettino* was the first magazine in Europe exclusively dedicated to the History of mathematics and physics. (Fiocca 2017).

1. fitting conditions for political, historical, philosophical and sociological speculations inside students movements;
2. actions aimed at dissemination of scientific culture;
3. a rich publishing of books dealing with history and philosophy of science.

### 3.1. *Fitting conditions*

Until the 90s, an invisible iron curtain and true borders divided the world into two blocks. Any kind of scientific communication between the two parts was very difficult and activities aimed at passing through the frontiers was suspicious. The establishment of CERN, strongly inspired among others by Edoardo Amaldi, had already successfully tried to break down the barriers between scientists by favoring basic research, without military outcomes. (Battimelli, Paoloni, 1998).

Farsighted was the vision of Gilberto Bernardini who, between the 1965 and the 1968, promoted the establishment of the European Physical Society (EPS), whose strategy was envisaged by CERN-based physicists as a response to U.S. hegemony in physics. The EPS was one of the very few specifically European international scientific organizations to transcend the Cold War political divide (Lalli 2021). On the history of the establishment of the EPS see also (Bevilacqua *et al.* 1993).

Several actions without military implications were taken by Scientific societies to overcome misunderstandings and suspicions among scientists. The Italian Physical Society organized a "Enrico Fermi School" on the Foundations of Quantum Mechanics and a School on the History of Twentieth Centrury Physics open to people coming from everywhere.

For a reconstruction of the contribution of physicists to the Foundations of Quantum Mechanics and, in general, to the History of Physics see (Baracca, Del Santo 2016), (Baracca *et al.* 2017), (Del Santo 2022). In the paper the authors reconstruct, based on the minutes of the Scientific Council, the contribution of the Italian Physical Society (SIF) to training of physicists in the field of the Foundations of Quantum Mechanics and the History of Science of the twentieth century. (D'Espagnat 1971)

In the 1972, when a war of liberation was underway in Vietnam, SIF organized a "Enrico Fermi School," for the first time dedicated to the History of Twentieth Centrury Physics instead of a frontier researches on contemporary physics (Weiner 1977). The School was attended by physicists, historians, philosophers, politicians (Weiner 1977 p. XI).

In Erice a "Advanced school of history of physics" directed by Giorgio Tabarroni was organized. I'm unable to give more details.

Also the Domus Galilæana, in the 1977, organized in Erice a "International School of History of Science" on the nature of scientific discovery, under the direction of Vincenzo Cappelletti and Mirko Grmek. Its aim was to provide contributions to topics straddling between philosophy and history of science. (Grmek *et al.* 2012)

From cultural and institutional point of view farsighted choices were the ones of the Domus Galilæana which, under the direction of L. Geymonat and V. Cappelletti, dedicated resources to the training of Historians of science and Historians of physics.

### 3.2. *Dissemination of scientific culture*

In the 1963, in view of the four hundredth anniversary of Galilei's birth (1564-1642), Bertold Brecht's "Vita di Galileo" (Galileo's life) was performed at the Piccolo Teatro in Milan. The version was the one staged in Berlin in the 1957. While in the original version of the 1938 Galilei was represented as a hero of compromise who bowed to power, but only to be able to continue his research, in the new version of the 1957, when the devastating effects of the bomb dropped in Hiroshima just when Brecht was in California had spread throughout the world, Galileo's abjuration was considered a betrayal of the ideals of the new science. Galileo, therefore, had been heroic in opposing the holy roman church but betrayer when he abjured. Consequences of Galilean behavior, according to Brecht, had been devastating; the link with the atomic bomb was obvious. But even to a philosopher of science, such as Geymonat, who claimed that the theory of knowledge in philosophy was nothing other than the theory of scientific knowledge, stressed

... the growing relevance of the problem raised by Brecht in the drama in question, namely the problem of the responsibility of scientific research in the ethical-philosophical-political field. (Geymonat 1963)

Galileo with his abjuration would have provided the first, very serious, example of disengagement of science.
The performance was accompanied by a series of initiatives aimed at spreading cultural values of science among public at large: a Galilean exhibition was staged in the theatre and a series of lectures was organized. It was the first times in Italy, liberated from fascist regime, that initiatives were planned to disseminate values of scientific culture. The problem of the responsibility of scientists, closely linked to the spread of anti-scientific conceptions, will become topical in the following decades.
In the 1968 the Club of Rome was founded by Aurelio Peccei and Alexander King and, in the 1972, *The limits of growth* was published in Italian (Meadows *et al.* 1972). The book challenged the thesis, supported by various traditional schools of economic thought, that only a program of modernization and industrialization would allow the progress of all mankind. The book and the topics proposed by the Club of Rome were not very successful, even among the movements that demanded freedom and equality. Its issues were branded as bad and erroneous science.
Since the '90s, the arguments of the Club of Rome have become increasingly topical and, a text published 50 years later, highlights how the predictions of the 1972 are coming true (Bardi, Pereira 2022).

*3.3.    Publishers' commitment*

The two Feltrinelli series of History of Science directed by Paolo Rossi and that one of Philosophy of Science directed by Ludovico Geymonat allowed the entry into the Italian debate of themes that elsewhere were of great topicality. Other publishing houses were also active. Below, in the footnote, I'll list some books that will be seminal in the following decades. It goes without saying that the list is by no means exhaustive and highlights only my preferences.[8]

---

[8] Thomas S. Kuhn *The Structure of Scientific Revolutions*; Rachel Carson published *Silent Spring* (Carson 1962); Charles Percy Snow *The Two Cultures* (Snow 1964). Ludovico Geymonat wrote the Preface to the Italian translation; Simone de Beauvoir *Le deuxième sexe* was published in Italian in the 1961 (first French edition 1949). The book was listed in the Index of forbidden books in the 1956; in the 1955 the Kinsey Reports: The Sexual Behavior of Man, and *The Sexual Behavior of Women* were published; Marcello Cini published "Il satellite della luna" (Cini 1969); Karl R. Popper's *The logic of the scientific discovery* was published in Italian (Popper 1971); Ludovico Geymonat *Storia del pensiero filosofico e scientifico* (Geymonat 1970-1996); Enrico

Some of the quoted books did not immediately become seminal. For example, Kuhn's book is not mentioned either in *Attualità del materialimo dialettico* or in *L'Ape e l'Architetto* which, in the second half of the 70s, animated the domestic debate around the nature of scientific knowledge.

Carson's text, and generally everything that had to do with the destruction of the natural environment, was rather neglected in the debate within the protest movements of the 70s and 80s. The same happened with the gender issues raised by de Beauvoir's book. But the time will tell.

## 4. Institutional framework

Increase of courses in History of physics in the 70s and 80s was not accompanied by recruit rules, but only by partial adjustments. Detailed papers by Paolo Rossi delineate various changes of Degree Courses in Physics. From it I'll enucleate the part that deals specifically with the History of physics and astronomy.

In the 1960 all physics courses were grouped in three sectors: "Physics education/teaching", "General Physics", "Applications of Physics". History of physics was fundamental in the sector "Physics education/teaching".[9] In the 1990 the sectors increased from three to nine: one of them was "Physics education/teaching and History of physics". Physics courses were compacted into four groups; History of Physics was listed in the group B01 ("General and Applied Physics"). In the 1994, B01 split into three sectors: one of this was B01C called "Physics education/teaching and History of physics" which in the 1999 became FIS/08. In B01C History of astronomy appears for the first time. (Rossi 2021, <https://osiris.df.unipi.it/~rossi/Fisici%20universita%201860-2010.pdf>, access date 11.11.2022>)

Listing History of physics among the physics courses was however a manifestation of interest of the physics community. But, if on the one hand the physicists had included the History of physics among the physical disciplines - and this was considered by many historians a recognition of efficacy of their research - on the other hand physicists were numerically prevalent in competition committees. Scholars in Physics Education/Teaching and in History of physics were often subjected to competitive judgments exercised by researchers who had little to do with research in Physics Education/teaching or History of physics (Bergia 1985, pp. 426-427).

A further difficulty of the historians in physics departments arose when the Minister established assessment criteria which included all physics sectors, included "Physics education/teaching and History of physics", among bibliometric sectors. Exceptions have been introduced over the years, but they have had the effect, in my view, of convincing physicists that research in history is a second-class research.

Some signals show that many physicists feel difficulty in adhering to a purely quantitative criterion but if the sector is managed according to rules that the Ministry has confirmed the exit routes are hard to travel.

I will address issues of quantitative assessment in the next paragraph.

---

Bellone, Ludovico Geymonat, Giulio Giorello and Silvano Tagliagambe *Attualità del materialismo dialettico* (Bellone *et al.* 1974); Giovanni Ciccotti, Marcello Cini, Michelangelo de Maria and Giovanni Jona Lasinio *L'ape e l'architetto.* (Ciccotti *et al.* 1976); *Criticism and the growth of knowledge* (Lakatos, Musgrave 1976).

[9] The decree "26 luglio 1960 n. 1692" stated that "The course of History of Physics must naturally contemplate the evolution of thought and physical theories and not simply the succession of facts and discoveries. The course can be introduced gradually, as appropriately qualified teachers are trained."

## 5.    The years '90

Since the end of the 80s the Italian university system has entered a state of continuous changes, which profoundly altered its features: selection of academic staff and regulation of careers, autonomy of universities, internal organization and relationship with other public and private subjects undergo profound transformations. (Colarusso, Giancola 2020) I will not go into details of these phenomena, but I will try to highlight the impact they have had on SF&SA.
New laws fell just when researchers in physics were involved in complicated changes in their role in world politics. I'll focus my attention on four points:
1. The role of the physicists on the international scene
2. Assessment of physics research
3. Emergence of new scientific disciplines
4. Historians of physics and astronomy in a changing world

*5.1 Physics and physicists on the international scene*

Starting from the 90s the interest of physicists for the history of their discipline and, in any case, for researches not strictly related to physics sectors, begun to decrease. Reasons are various.
"The development of the nuclear bomb brings up conflicting feelings." So Robert Oppenheimer, in 2015, commented the anniversary of the first test of the atomic bomb in July 1945 (Editorial of Nature physics 2015). In a paper of *Physics Today* the physicists' dilemma about nuclear weapons are clearly described:

The physics community has a special relationship to nuclear weapons policy. Physicists invented and refined nuclear weapons and historically have made major contributions to efforts to limit the dangers they pose. (Fetter et al. 2018, p. 39)

In a bipolar world, dominated by two great powers, holder of thousands of nuclear devices, physicists on both sides have been involved to give their contribution to armaments. Moreover, they have been the guarantors and controllers, with their skills, that the balance based on the terror of a nuclear disaster was not altered. At the same time physicists have animated various national and international organizations that have stemmed proliferation of nuclear weapons: Pugwash, Physicists coalition for nuclear threat reduction of the American Physical Society (APS); Italian Union of Scientists for Disarmament (USPID). (Clavarino 2021), (Greco 2017).
Another theme that impacted the role of physicists in the international arena was what the Club of Rome had already stated half a century ago. Since the seventies of the last century, a growing awareness of the limited nature resources, and ineffectiveness of science to replace them with others resulting from scientific and technological research, had been spreading. Physicists have been wrongly believed to be responsible for this impotence.
One consequence, presumably unforeseen by the same physicists who had opposed the war and civilian developments of nuclear energy, was the spread of anti-scientific positions that questioned the way in which science and technology addressed unresolved problems. Those who didn't give any credit to science denied that values underlying scientific knowledge could be an important component of modern societies.

### *5.2 Quantitative assessment of researches*

Publications in the physics field since the last decades of the last century had grown at a very high rate. To safeguard the values intrinsic to science, scientists, including physicists, have imposed external quantitative criteria.
The effect, probably not expected, was alien to scientific tradition: personal judgments have been replaced by formal evaluations, analysis of texts had been replaced by numerical counts, quality by quantity. (Renn 2017)
Specialized and sectoral scientific results contributed to the fragmentation of science and physics, as well as to the exclusion of research that goes beyond disciplinary boundaries. No physicist, from a certain point on, has been able to assess the quality of publications in areas in which physics had differentiated. (Renn 2017, p. 574).
Standardization for physics papers didn't fit the assessment of physics history papers. Not to mention the essays that are not even considered.
To date, a solution has not been found to the right request of physicists for a less subjective method in scientific papers assessment and, at the same time, to historians' call to be judged with criteria appropriate to the type of research carried out. If the weight of historians of physics department is so small and courses of history are so few the two requests are difficult to reconcile.

### *5.3 Emergence of new scientific disciplines*

In the last decades of the '900 needs emerged that only a science without epistemological barriers and without national borders could face: climate change, nuclear proliferation, clean energy, sustainable economic and demographic development, adequate world distribution of food, gender barriers.
The topics were not new: just think at *Silent Spring* by Rachel Carson, or the birth of the Club of Rome, or de Beauvoir's book.
Starting from the last two decades of the last century these themes have reached a level of diffusivity and urgency unimaginable. Even in the protest movements of the 60s and 70s these issues were underestimated except for the peaceful use of nuclear energy.
In connection with new needs some ancillary sciences have evolved into autonomous sciences: meteorology, seismology, biophysics, earth sciences, space sciences, computer science. In connection with these changes the history of science itself had to reorient its field of investigation. The history of physics was no longer a central topic in the history of science and Physics no longer defined the agenda of historians, as it was the case for Kuhn, Klein and Dijksterhuis (Crease *et al.* 2020). Many of the new sciences use methods and results elaborated in physical sciences. The new sciences have been considered a threat to physics cultural influence into scientific world.

### *5.4 Historians of physics and astronomy facing new challenges*

The Society of Historians of Physics and Astronomy (SISFA) was established on March 23, 1999. The first President was Pasquale Tucci, who have been the first to get a chair in History of physics requested by a physics department through a competition announcement. The Society inherited the legacy of GNSF. The CNR, in analogy to what happened to other coordination groups, had decided to cancel the National Coordination Groups. GNSF became a Commission; for the first time there were also Historians of astronomy.

The Society ensured continuity with the GNSF as far as a competition for new university positions was concerned: just as Tagliaferri had wanted since the establishment of the GNSF, the Society did not enter in any way into the dynamics of the competitions.

The Society included Historians of astronomy, an obvious choice from the point of view of method and affinity of the contents of the research.

A strong element of continuity, which survived to all fluctuating events of SISFA, was the annual Conference of GNSF and of SISFA: the Proceedings hold a very rich documentation. It has been a shared element over the years to avoid filters on participation and presentation of contributions both in the Conferences and in Proceedings.

A good source of funding was reached when the CNR, in 1995, approved the "Cultural Heritage" project. Giorgio Dragoni (Dragoni 1993) had collected an impressive documentation of historical scientific instruments scattered among several Italian research institutions, for the most part, in extreme danger of dispersion and destruction. The subproject 5 concerned "Museology and Museography" and one of the themes - 5.4 - was "Cataloguing, conservation and restoration of cultural heritage".

Another source of funding came from the Law for the Dissemination of Scientific Culture of the Ministry of University and Research.

Compared to massive changes, reaction of the historians had been to analyze the changes with the methodology and tools of the historical sciences, avoiding to follow physicists on the hypertrophic accumulation of papers:

- many historians of physics have dealt with research topics concerning the history of nineteenth - and twentieth-century physics. On such historical issues it is easier to communicate with physicists;
- many initiatives have been carried out to disseminate scientific culture in relation to the growing skepticism of public opinion towards physics, arbitrarily linked to environmental disasters;
- connection between teaching and history of physics has been a theme of research and initiatives pursued with undoubted results in some departments;
- enhancement of historical-scientific heritage which can become a museum, but also a physical and immaterial context for activities of dissemination of scientific culture.

Over the years, historians' awareness of the nature of scientific knowledge has changed: the paradigm of progress that had accompanied science since its origins had impressed many historians of science. But over the years the idea that the history of science should be considered as cultural history or social anthropology had become increasingly stronger: science could no longer offer a model of rationality to be applied to all domains of human life. On the other hand, it should not be overlooked that it is *a special form of knowledge* that is affected by the historical context. (Renn 2020, pp. 13-15)

While physicists have continued to have many doubts, even legitimate, about quality of historical researches, they have been much more sensitive about the initiatives to spread scientific culture and about the connection between history and teaching. Probably because they saw in them a way to increase appeal of physics among the students of the various pre-university schools.

## 6. Conclusions

For many historians of physics the last 50 years have been exciting, although objective difficulties and real errors of the various subjects who have had roles of responsibility have not been lacking. Since the establishment of GNSF, what mattered then and that, in my opinion, in the future will have more and more relevance, are papers published in the international journals of History, in the Proceedings of conferences, in Books and in other publications. Researches and Publications have to be a priority over all other activities.

For sure administrative and legislative framework must change: our objective is to implement all possible strategies so that historical researches and teaching in degree courses in physics have to be considered in the same way as other researches and teaching in physics, purifying them of trappings and quantitative artifices. Every degree course of scientific nature should include courses in History of the discipline, Epistemology and Sociology, as Aldo Mieli had hoped for almost a century ago.

A historical, epistemological, sociological background can help future citizens to relate to the most advanced and widespread aspects of modernity: without illusions of uninterrupted progress but also without unjustified suspicions about the capacity of science to make more livable the world in which we live and work: science, and not only its technological applications, is a fundamental stage in human civilization.

## References

Amaldi, E. (1973). "The unity of physics". *Physics Today.* 26(9), pp.23-29.

Amaldi, E. (1977). "Personal Notes on Neutron Work in Rome in the 30s" in Weiner, C. (ed) (1977). *Proceedings of the International School of Physics "Enrico Fermi". History of Twentieth Century Physics 1972*. Bologna and London: SIF and Academic Press, Vol. 57, pp. 294 - 325.

Baracca, A.; Bergia, S. and Del Santo, F. (2017). "The origins of the research on the foundations of quantum mechanics (and other critical activities) in Italy during the 1970s. *Studies in History and Philosophy of Science Part B: Studies in History and Philosophy of Modern Physics* 57, pp. 66-79.

Bardi, U.; Pereira, C. A. (eds) (2022). *Limits and Beyond. 50 years on The Limits to Growth, what did we learn and what's next?* Exapt Press.

Battimelli, G.; Paoloni, G. (eds) (1998). *20th Century Physics: Essays and Recollections, a Selection of Historical Writings*. Singapore; River Edge, NJ: World Scientific.

Battimelli, G.; De Maria, M.; La Rana, A. (eds) (2022). *Da via Panisperna all'America.* Roma: Editori Riuniti.

Bellone, E.; Geymonat, L.; Giorello, G.; Tagliagambe, S. (1974). *Attualità del materialismo dialettico* Roma: Editori Riuniti.

Bergia, S. (1985). "Ricerche di storia della fisica in Italia" in D'Agostino, Salvo; Petruccioli, Sandro (eds.). *Atti del V Congresso Nazionale di Storia della Fisica*. Roma: Istituto Poligrafico e Zecca dello Stato, pp. 425-440.

Bergia, S.; Del Santo, F. (2016). "The origins of the research on the foundations of quantum mechanics (and other critical activities) in Italy during the 1970s". Revised in 2016. <https://arxiv.org/pdf/1608.05197.pdf>, access date: 03.06.2021

Bevilacqua, F.; Esposito, S. (2021). "SISFA: 40 years of history of physics in Italy". *Il Nuovo Saggiatore. Bollettino della Società Italiana di Fisica.* NS 37(1-2), pp. 39-50.

Bevilacqua, F.; Giannetto, E.; Tagliaferri, G. (1993). "Europe in 1965-1968" *Europhysics News* 24, pp. 115-117.

Borrelli, A.; Schettino, E. (2005). "La prima cattedra di storia della fisica in Italia: un'occasione mancata" *Scienza & Politica* 33, pp. 75-94.

Braccini, S.; Ereditato, A.; Scampoli, P. (eds) (2012). *Edoardo Amaldi: The Adventurous Life of Friedrich Georg Houtermans, Physicist (1903 - 1966).* Berlin: Springer.

Carson, R. (1962). *Silent spring* Boston: Houghton Mifflin Company. Edizione italiana: Carson R. (1963) *Primavera silenziosa* Milano: Feltrinelli.

Ciccotti, G; Cini, M.; de Maria, M. e Jona Lasinio, G. (1976). *L'ape e l'architetto* Milano: Feltrinelli.

Clavarino, L. (2021). "Italian Physicists and the Bomb: Edoardo Amaldi's Network for Arms Control and Peace during the Cold War" *Journal of contemporary history* 56(3), pp. 665-692.

Colarusso, S.; Giancola, O. (2020). *Università e nuove forme di valutazione. Strategie individuali, produzione scientifica, effetti istituzionali* Roma: Sapienza Università Editrice.

D'Espagnat, B. (ed) (1971). *Proceedings of the International School of Physics "Enrico Fermi". Foundations of Quantum Mechanics 1970.* Bologna and London: SIF and Academic Press, Course n. 49.

Crease, R.P.; Martin, J.D. & Staley, R. (eds) (2020). Editorial "Recentering the History of Physics". *Physics in Perspective* 22(1), pp. 1-2.

de Beauvoir, S. (1949). *Le deuxième sexe* Paris: Gallimard. Edizione italiana: Beauvoir, S. (1961). *Il secondo sesso* Milano: il Saggiatore.

Del Santo, F. (2022). "The Foundations of Quantum mechanics in post-war Italy's cultural context" in Freire, O. jr; Bacciagaluppi, G.; Darrigol, O.; Hartz, T., Joas, Christian, Kojevnikov, A. and Pessoa, O. jr (eds) (2022). *The Oxford handbook of the History of quantum interpretations* Oxford: Oxford University Press, pp. 641-666. <https://doi.org/10.1093/oxfordhb/9780198844495.013.0026, access date 7 July 2022>

Drago, A. (1994). "Riflessioni sui corsi di Storia della fisica" in *Venti anni di didattica universitaria di Storia della fisica* in Bevilacqua, F. (ed) (1994). *Atti del XII Congresso nazionale di Storia della fisica* Milano: Gruppo Nazionale di coordinamento per la Storia della fisica, pp. 295-296.

Dragoni, G. (1993). "La strumentazione storico scientifica: sintetica rassegna nazionale delle attività del GNSF" in Bitelli Masetti, L. (ed) (1993) *Restauro di strumenti e materiali: scienza, musica, etnografia* Bologna: Istituto per i beni artistici, culturali e naturali della Regione Emilia-Romagna, pp. 13-49.

Duhem, P. (1913). *Le système du monde. Histoire des doctrines cosmologiques de Platon à Copernic*. Paris: Hermann.

Editorial (2015). "Physics, physicists and the bomb" *Nature Physics*, 11(3), p. 201.

Favaro, A. (1889). "Il Bullettino di Bibliografia e di Storia delle Scienze Matematiche e Fisiche pubblicato da D.B. Boncompagni (1868–1887)" *Bibliotheca Mathematica* Neue Folge 3, 4, pp. 109-112.

Fetter, S.; Garwin, R.L.; Hippel, F. (2018). "Nuclear weapons dangers and policy options" *Physics Today* 71(4), pp. 32-39.

Fiocca, A. (2017). "The Bullettino di Bibliografia e di Storia delle Scienze Matematiche e Fisiche (1868–1887), an example of the internationalisation of Research" *Historia Mathematica* 44, pp.1-30.

Gariboldi, L. (2022a). "Giovanni Polvani and the Institute of Physics before the Second World War" in *The Milan Institute of physics.* Cham: Springer International Publishing, pp. 55-85.

Geymonat, L. (1963). "Il Galileo della storia e il Galileo di Brecht" *L'Unità*, 2 aprile 1963.

Geymonat, L. (ed) (1971-1996). *Storia del pensiero filosofico e scientifico* 11 vols. Milano: Garzanti.

Greco, P. (2017). *Fisica per la pace: tra scienza e impegno civile.* Roma: Carocci.

Ienna, G. (2022). "Le origini politico-istituzionali del Gruppo Nazionale di Storia della Fisica" in Zanini, V.; Naddeo, A. and Bònoli, F. (eds) (2022). *Atti del XLI Convegno annuale della Società Italiana degli Storici della Fisica e dell'Astronomia. Arezzo, 6-9 Settembre 2021.* Pisa: Pisa University Press, pp. 15-22

Kaiser, D. (2012). "In retrospect: The structure of scientific revolutions" *Nature* 2012 484, pp. 164-165.

Kinsey, A.C.; Pomeroy, W.B; Martin, C.E. (1948). *Sexual behavior in the human male* Philadelphia and London: Saunders Company. Edizione italiana: Kinsey, A.C.; Pomeroy, W.B; Martin, C.E.; Musatti, C.L. (1955). *Il comportamento sessuale dell'uomo* Milano: Bompiani.

Kinsey, A.C.; Pomeroy, W.B; Martin, C.E.; Gebhard, P.H. (1953). *Sexual behavior in the human female* Philadelphia and London: Saunders Company. Edizione italiana: Kinsey, A.C.; Pomeroy, W.B; Martin, C.E.; Gebhard, P.H. (1955). *Il comportamento sessuale della donna* Milano: Bompiani.

Lakatos, I. e Musgrave A. (eds) (1974). *Critica e crescita della conoscenza* Milano: Feltrinelli. Prima edizione Lakatos, I. e Musgrave A. (eds) (1970). *Criticism and the growth of knowledge* Cambridge: Cambridge University Press.

Lalli, R. (2021). "Crafting Europe from CERN to Dubna: Physics as diplomacy in the foundation of the European Physical Society" *Centaurus* 63, pp. 103-131.

Mandrino, A.; Tagliaferri, G.; Tucci, P. (ed) (1994). *Un viaggio in Europa nel 1786. Diario di Barnaba Oriani astronomo Milanese* Firenze: Leo Olschki editore, 1994.

Meadows, D.H.; Meadows, D.L.; Randers J.; Behrens III, W.W. (1972). *The limits of growth* Washington: Potomac Associates Book. In Italian: *I limiti dello sviluppo* Milano: Edizioni scientifiche e tecniche Mondadori.

Mieli, A. (1916). *La storia della scienza in Italia. Saggio di bibliografia di storia della scienza* Firenze: Libreria della voce.

Miotto, E.; Tagliaferri, G.; Tucci, P. (1989). *La strumentazione nella storia dell'Osservatorio astronomico di Brera* Milano: Unicopli Editore.

Pancaldi, G. (1980). "The History and Social Studies of Science in Italy" *Social Studies of Science* 10(3), pp. 351-374.

Pancaldi, G. (2020). "Italy" in Slotten, Hugh Richard; Numbers, Ronald L.; Livingstone, David N. (eds.) (2020). *The Cambridge History of Science. Modern Science in National, Transnational, and Global Context* Cambridge: Cambridge University Press, Vol 8, pp. 435-454.

Pancaldi, G. (2020). "ITALY" in Slotten, Hugh Richard; Numbers, Ronald L.; Livingstone, David N. (eds) (2020). *The Cambridge History of Science. Modern Science in National, Transnational, and Global Context* Cambridge: Cambridge University Press, Vol 8, pp. 345-360.

Pogliano, C. (1983). "Aldo Mieli, Storico della Scienza (1879-1950)" *Belfagor* 38(5), pp. 537-557.

Polvani, G. (1942). *Alessandro Volta.* Pisa: Domus Galilæana.

Popper, K.R. (1971). *Logica della scoperta scientifica. Il carattere autocorrettivo della scienza* Torino: Einaudi. Traduzione di Popper, K.R. (1935). *Logik der Forschung* Vien: Julius Springer.
Renn, J. (2017). *The Globalization of Knowledge in History* Berlin: Edition Open Access.
Renn, J. (2020). *The Evolution of Knowledge.* Princeton: Princeton University Press.
Rossi, P. (2015). "I fisici nel sistema universitario italiano 1980-2015" *Il Nuovo Saggiatore. Bollettino della Società Italiana di Fisica* NS 2015 31(3-4), pp. 80-87.
Robotti, N. (2002). "Guido Tagliaferri docente di storia della fisica" *Giornale di Fisica* XLIII(3), pp. 171-175.
Rossi, P. (2021). "I Fisici nel sistema universitario italiano (1860-2010)" <https://osiris.df.unipi.it/~rossi/Fisici%20universita%201860-2010.pdf, access date 11.09.2022>).
Rubbia, C. (1991). "Edoardo Amaldi. 5 September 1908-5 December 1989". *Biographical Memoirs of Fellows of the Royal Society* 37, pp. 1-31.
Salvetti, C. (2002). "Lettera a Tagliaferri" *Giornale di fisica* XLIII(3), pp. 177-180.
Salvini, G.; Tucci, P. (2000). "Guido Tagliaferri, fisico, storico, umanista" *Il Nuovo Saggiatore. Bollettino della Società Italiana di Fisica* NS 16 (5-6), pp. 65-70.
Sarton, G. (1913). "L'Histoire de la Science" *ISIS* 1, pp. 3-46.
Sarton, G. (1927). *Introduction to the History of science* Baltimore: Williams & Wilkins company, 3 vols.
Tagliaferri, G. (1985). *Storia della fisica quantistica: Dalle origini alla meccanica ondulatoria* Milano: Franco Angeli.
Tagliaferri, G. (1994). "L'insegnamento della Storia della fisica alla Facoltà di Scienze MF&N dell'Università di Milano" in Drago, A. (ed) "Venti anni di didattica universitaria di Storia della fisica tenutosi a Napoli il 1989" in Bevilacqua, F. (ed) (1994). *Atti del XII Congresso nazionale di Storia della fisica* Milano: Gruppo Nazionale di coordinamento per la Storia della fisica, pp. 373-377.
Tagliaferri, G.; Tucci, P. (1999). "Carlini and Plana on the theory of the Moon and their dispute with Laplace" *Annals of Science* 56(3), pp. 221-270.
Torrini, M. (1988). "Observations on the History of Science in Italy". *The British Journal for the History of Science* 21, pp. 427-446.
Tucci, P. (2007). "Il Museo Astronomico e l'Orto Botanico di Brera in Milano" *Annali di storia delle Università italiane* 2007, pp. 251-259.
Weiner, C. (ed) (1977). *Proceedings of the International School of Physics "Enrico Fermi". History of Twentieth Century Physics 1972*. Bologna and London: SIF and Academic Press, Vol. 57.